\newcommand{\eg}[0]{{\it e.g.}}
\newcommand{\ie}[0]{{\it i.e.}}
\newcommand{\etal}[0]{{\it et al.}}

\documentclass[12pt]{article}

\usepackage{amssymb,fullpage,url,times,hanging,footmisc}

\begin{document}

\pagenumbering{roman}
\renewcommand{\thefootnote}{\fnsymbol{footnote}}

\begin{center}
{\Huge Training the Next Generation of Astronomers%
\footnote{This document was composed primarily at the University of
  California at Berkeley. As such, the majority of authors are
  associated with this institution. However, while the opinions
  expressed here are endorsed by the authors, they do not reflect
  those of the institution as whole. The authors consist of 15
  graduate students, 15 postdocs, and one permanent research staff.}}
\end{center}

\renewcommand{\thefootnote}{\arabic{footnote}}

\begin{center}
  {\bf Primary Author:} Peter K.~G. Williams$^{1,2}$
  \\
  {\bf Coauthors:}
  Eric Huff\footnotemark[1], 
  Holly L. Maness\footnotemark[1], 
  Maryam Modjaz\footnotemark[1], 
  Kristen L. Shapiro\footnotemark[1], 
  Jeffrey M. Silverman\footnotemark[1], 
  Linda Strubbe\footnotemark[1];
  Betsey Adams\footnotemark[3],
  Katherine Alatalo\footnotemark[1],
  Kuenley Chiu\footnotemark[4],
  Mark Claire\footnotemark[5],
  Bethany Cobb\footnotemark[1],
  Kelle Cruz\footnotemark[4],
  Louis-Benoit Desroches\footnotemark[1],
  Melissa Enoch\footnotemark[1],
  Chat Hull\footnotemark[1],
  Hannah Jang-Condell$^{6,7}$,
  Casey Law\footnotemark[1],
  Nicholas McConnell\footnotemark[1],
  Rowin Meijerink\footnotemark[8],
  Stella Offner\footnotemark[9],
  John K. Parejko\footnotemark[10],
  Jonathan Pober\footnotemark[1],
  Klaus Pontoppidan\footnotemark[8],
  Dovi Poznanski$^{1,11}$,
  Anil Seth\footnotemark[12],
  Steven Stahler\footnotemark[1],
  Lucianne Walkowicz\footnotemark[1],
  Andrew A. West\footnotemark[13],
  Andrew Wetzel\footnotemark[1],
  David Whysong\footnotemark[1]
\end{center}
\footnotetext[1]{Department of Astronomy, University of California at
  Berkeley, Berkeley, CA 94720}
\footnotetext[2]{Tel.\ 510 642 5189; email \url{pwilliams@astro.berkeley.edu}}
\footnotetext[3]{Astronomy Department, Cornell University, 
  Ithaca, NY 14853-6801}
\footnotetext[4]{Astronomy Department,
  California Institute of Technology, Pasadena, CA 91125}
\footnotetext[5]{Department of Astronomy, University of Washington,
  Box 351580, Seattle, WA 98195}
\footnotetext[6]{Department of Astronomy, University of Maryland,
  College Park, MD 20742}
\footnotetext[7]{NASA Goddard Space Flight Center, Code 685, 
  Greenbelt, MD 20771}
\footnotetext[8]{Department of Geological and Planetary Sciences,
  California Institute of Technology, Pasadena, CA 91125}
\footnotetext[9]{Department of Physics, University of California at
  Berkeley, Berkeley, CA 94720}
\footnotetext[10]{Department of Physics, Drexel University, 
  Philadelphia, PA 19104}
\footnotetext[11]{Lawrence Berkeley National Laboratory, 1 Cyclotron
  Road, Berkeley, CA 94720}
\footnotetext[12]{Harvard-Smithsonian Center for Astrophysics, 
  60 Garden Street, Cambridge, MA 02138}
\footnotetext[13]{Department of Physics, Massachusetts Institute of
  Technology, Cambridge, MA 02139}

\begin{center}
\section*{Abstract}
\end{center}
While both society and astronomy have evolved greatly over the past
fifty years, the academic institutions and incentives that shape our
field have remained largely stagnant.  As a result, the astronomical
community is faced with several major challenges, including: (1) the
training that we provide does not align with the skills that future
astronomers will need, (2) the postdoctoral phase is becoming
increasingly demanding and demoralizing, and (3) our jobs are
increasingly unfriendly to families with children. Solving these
problems will require conscious engineering of our profession.
Fortunately, this Decadal Review offers the opportunity to revise
outmoded practices to be more effective and equitable. The highest
priority of the Subcommittee on the State of the Profession should be
to recommend {\it specific, funded} activities that will ensure the
field meets the challenges we describe.

 
\clearpage
\pagenumbering{arabic}
\setcounter{page}{1}
\setcounter{footnote}{0}

\section{Introduction}

The collective self-examination made possible in astronomy by the
Decadal Review process is a prime opportunity to engineer the
incentives and institutions that shape our profession.  We need not
simply accept as inevitable the institutional framework that we've
inherited. 

The standard trajectory of the American academic career has been
essentially fixed since the mid-20th century, when postdoc
appointments started becoming common [1].  However, the practice of
astronomy has changed since the 1950's: we now deal with increasingly
enormous telescopes, collaborations, and data sets. The lack of
similar evolution in graduate training has resulted in
Ph.D. recipients who are no longer optimally trained for the skills
and new positions required by modern astronomical research. In \S2, we
discuss these issues in detail and identify inefficiencies in the
current structure for funding and training new professionals.

The practical structure of academic astronomy has also changed
significantly.  While the Ph.D. overproduction rate compared to
faculty spots has remained approximately steady over the past two
decades, there were many fewer postdoc positions in the past [2]. In
recent years, however, increased federal funding has led to a boom in
graduate student and postdoc positions without a concomitant expansion
in the number of permanent faculty positions [3]. In \S3, we explore
the implications of these realities in our field.

Additionally, the past half-century has witnessed a dramatic change in
the workforce. The opportunity to gain the contributions of many
excellent astronomers is currently missed. In \S4, we consider only
one example of this phenomenon: how the resulting rise in the demand
for ``child-friendly'' careers has been borne by our field. Progress
to date has been minimal, and the evidence is that this situation
systematically selects against women.

There is thus a good case to be made that the institutional framework
of academic astronomy is suboptimal and disserves both the
practitioners of astronomy and the public that ultimately funds it.
Fortunately, this is an issue that we, the astronomical community, can
solve. We argue that the Subcommittee on the State of the Profession
(SSP) should direct its attention toward improving the structure of
our instutional framework.  We outline below the evidence for a few
outstanding problems, describe their costs to the community, and
provide some suggestions that we hope will prove useful to those
charged with charting the course of astronomy over the next decade.

\section{Allocation of Training Resources}

The training of professional astronomers is integral to the discussion
of what science will be done in the coming ten years. In this section,
we discuss the current state of astrophysics training and propose that
adaptations to this process will be necessary in order to best use
financial resources and personnel to produce the best science.

\subsection{The Cost, Training, and Employment of Astronomy Ph.D.s}

Significant financial resources are currently invested in the training
of the next generation of astronomers.  A typical astronomy Ph.D.
candidate may earn \$20K annually, with an additional \$20K spent by
the PI or department to pay tuition and university fees.  Assuming
that the average student spends 5 years in graduate school and
requires an additional \$50K over the course of this time to support
equipment, travel, and miscellaneous expenses, every graduate student
will cost the profession a quarter of a million dollars.  Each year in
the past decade, there have been $\gtrsim$170 astronomy Ph.D.s awarded
in the United States [4, 5], {\it for a total of over \$43M spent
  annually to produce new professionals} (see also the Seth
\etal\ Position Paper on ``Employment \& Funding in Astronomy'').
This is almost quadruple the annual operations budget of the Keck
Observatory [6].

Graduate students are typically funded through research grants and are
typically expected to devote the vast majority of their time to pure
research.  This policy is even expressly stated in some graduate
student guides (\eg\ [7]) and is pervasive in the professional
culture.  Few programs offer any incentives to broaden coursework
beyond astronomy and physics to include computer science, engineering,
public policy, business, or education.  While many astronomy graduate
programs mandate that students spend one or more semesters as teaching
assistants, training in teaching skills is generally minimal, although
there are laudable exceptions [8, 9].  Development of the skills
needed for teaching at the non-university level and public outreach is
typically absent from the curriculum.

There is only room in the field for $\sim$50\% of Ph.D. recipients to
become faculty at colleges and universities [3].  In fact,
self-reporting of the careers of 651 astronomy Ph.D. recipients from
1980-2000 at eight universities reveals that 34\% currently hold
tenure-track faculty positions at research universities\footnote{From
  the websites of astronomy departments at Harvard University, the
  Ohio State University, Princeton University, UC Berkeley, UCLA, UC
  Santa Cruz, UT Austin, and the University of Virginia.  Note that
  these departments are highly-ranked and likely have above-average
  tenure-track placement rates compared to the field as a whole. Data
  available in ASCII format online at
  \url{http://astro.berkeley.edu/~pkwill/ay2010_training/careers.txt}.}.
The remaining two-thirds of Ph.D. astronomers are employed at teaching
colleges as tenure-track faculty (10\%), at K-12 schools as educators
or elsewhere as education researchers (2\%), at observatories and
national laboratories as permanent support/research staff (38\%), and
within business and industry (17\%). That is, almost one third of
Ph.D. recipients are not primarily employed in research. 

An informal survey of UC Berkeley faculty indicates that they spend
approximately 25\% of their work time on research (excluding student
and postdoc interaction), 25\% on teaching, 19\% on administrative
duties (including committee participation and large-scale project
management), 14\% on advising students and postdocs, 12\% on securing
funding (for personal research as well as
observatories/organizations/departments as a whole), and 5\% on public
outreach\footnote{Data available in ASCII format online at
\url{http://astro.berkeley.edu/~pkwill/ay2010_training/factime.txt}.}. 
Despite the small sample size we feel that it is fair to say
that {\bf even ``research university'' faculty spend the majority of
  their time on activities other than research}.

Overall, we find that {\bf training is a significant expenditure in
  the field of astronomy} and that {\bf a majority of astronomy
  Ph.D. recipients spend a significant fraction of their time on
  activities other than research}.

\subsection{The Needs of the Field Now and in the Coming Decade}

Although the critical problem-solving skills obtained via research
training are unarguably used by all astronomy Ph.D.s, regardless of
career, there are other, equally valuable skills that these Ph.D.s
will need that are not currently included in graduate
training. This mismatch between the training of young professionals
and the skills required in their future employment will only continue
to grow in the coming decade.  The increasing size and scope of
projects in astronomy (\eg\ Keck, VLA, ALMA, TMT, supercomputing and
data management facilities) are creating increasingly large
collaborations in which diverse skills are needed [10].

Obviously, the success of these projects depends critically on the
$\sim$$\frac{1}{3}$ of Ph.D.s who become faculty and PIs. Such success
is, however, dictated by much more than the PI's ability to do excellent
science. Lead scientists within collaborations must also excel at
managing funding and personnel, communicating science needs and
results to the public (including the general public, government
agencies, private investors, and industrial partners), and instructing
and mentoring junior members of the group. Many of the requisite
skills for these tasks are, in general, completely ignored during
graduate and postdoctoral training and developed only later through
trial and error. The lack of these skills in PIs can result at best in
inefficiencies and at worst in misuse of funding and failure of strong
scientific programs.
 
About two-fifths of Ph.D. astronomers find employment in permanent
support or research staff positions, which require skills in areas
such as data management and the construction, operation, and
maintenance of hardware and software. This proportion will likely
increase with the increasing scope of planned facilities. Despite this
fact, Ph.D. programs generally do not have formal structures that
encourage candidates to develop sophisticated programming or
engineering skills.

Even more critically, the funding of the field and influx of talented
individuals into it rely on education and outreach at all levels.
This is the explicit career of 11\% of Ph.D.  astronomers (K-12
educators, education researchers, and professors at teaching
colleges), but it is also an important role of university faculty
($\sim$30\% of whose time is devoted to teaching and public outreach)
and, to varying extents, of support astronomers.  The importance of
this part of astronomy cannot be understated: in the notable case of
the Hubble Space Telescope (HST), public enthusiasm for the field
directly led to the continuation of NASA support and congressional
funding that would have otherwise been cut [11]. Nevertheless,
education and outreach typically have minimal roles in astronomy
training (\S2.1).

Finally, $\sim$$\frac{1}{6}$ of Ph.D. astronomers leave the field
entirely. While, as argued by the Seth \etal\ Position Paper on
``Employment \& Funding in Astronomy'', there are not enough permanent
positions in the field for all astronomy Ph.D.s, there is no reason to
believe that those who leave are uniformly less excellent astronomers
than those who stay. {\bf The training that Ph.D.s receive creates
  expectations about the profession and signals what it values. The
  mismatch between that training and actual employment opportunities
  may drive talented young scientists to leave the profession.}

\subsection{The Role of the Decadal Review}

The above breakdown of job outcomes is a statistical reality. Graduate
mentors need to both support and provide training for a number of
possible employment opportunities. We suggest that the SSP investigate
ways in which funding structures and associated directives to
universities can be altered to support this realignment. In
particular, {\bf we suggest the following}:

\parindent 0.5in

{\bf (A)} That the definition of a career in astronomy be broadened to
include the true assortment of potential careers in astronomy that
Ph.D.s eventually have; that this be assessed via rigorous tracking of
the employment statistics of all Ph.D. recipients from graduate school
through postdoc positions and culminating only when permanent
positions are attained; and that updated statistics be regularly
disseminated within the academic community. This is also one of the main
points of the Seth \etal\ Position Paper on ``Employment \& Funding in
Astronomy''. 

{\bf (B)} That the training given at universities granting astronomy Ph.D.s
reflect this paradigm shift, such that graduate students are trained
for the jobs they will eventually hold.

{\bf (C)} That communication and leadership skills be emphasized in a
meaningful and substantial way in Ph.D. programs, helping the next
generation to garner support from non-scientists, lead successful
scientific projects and collaborations, and attract, educate, and
mentor future generations of scientists.

\parindent 0.25in
 
Creative thinking will obviously be necessary to effect such a change
in the astronomy education system at the graduate level.  Here we list
some means through which this might be achieved, which we intend not
to be exhaustive but rather to initiate discussion: new Ph.D.
programs could be funded to provide the breadth of knowledge and
specialization now required by many careers in astronomy, \eg,
joint programs between astronomy and computer science, engineering,
business, public policy, or education (dual Ph.D., Ph.D./masters,
Ph.D. with a ``minor''); federally-funded astronomy Ph.D. students
could be required to spend a semester away from research in, \eg,
government agencies, student teaching positions, or internships within
industry; quantifiable mentoring, teaching, and outreach requirements
could be attached to federal grants (expanding on the latest NSF
Proposal and Award Policies and Procedures Guide%
\footnote{\url{http://www.nsf.gov/pubs/policydocs/pappguide/nsf09_1/index.jsp}}%
) such that young professionals are required to devote part of their
time to improving their leadership and communication skills. Most
likely, a combination of many changes and initiatives will be needed
to ensure that the training of young astronomers is best-suited to the
positions that will need to be filled in the next ten years and
beyond.

\section{Postdocs}

The structure of the traditional academic career has not kept up with
the realities of the modern university. The postdoctoral position,
once a short stopping point between Ph.D. and the tenure track, has
evolved into a substantial phase of the academic career, with
recipients holding 2--3 postdoc positions (\ie, 4--9 years) until a
permanent job is obtained [3]. Along with the increasing duration of
this phase, it is also becoming increasingly and unnecessarily
demanding and demoralizing.

\subsection{The Postdoctoral Experience}

Three fundamental factors are responsible for the transformation of
the character of the postdoc phase: (1) the boom in Ph.D.s granted,
(2) the lack of a similar expansion in permanent academic positions,
and, importantly, (3) the failure of most Ph.D. training programs to
adapt to this discrepancy, as described in \S2. The interaction of
these factors results in a situation where there are many people
competing for few spots. While intense competition for prestigious
jobs is natural, the incentives of the field encourage maximization of
the amount of work extracted from trained astronomers, \ie, the
attrition of Ph.D.s out of the running for those jobs as late as
possible. This situation does not select for higher-quality faculty
--- it merely places unnecessary burdens on those who do not end up
attaining faculty jobs. The explicit recommendation of the previous
Decadal Review committee to increase federal funding for postdoctoral
fellowships has played a role in this effect [12, p.\ 198].

The consequences of this situation are acute. {\bf Current postdocs
  endure a period of intense competition, prolonged job insecurity,
  multiple relocations with little or no choice in destination, and
  the prospect of a forced late-stage (mid-30's) career change.} As we
discuss in the following section, the implications for those who also
wish to start a family are particularly dire. This situation will not
change as long as the three factors described above hold. Because our
field is vibrant and competitive and academic jobs are very attractive
to many, it would be incorrect to suggest that the difficulty of the
postdoc phase will lead to unfilled tenure-track positions. However,
the simple fact is that {\it we can do better}.

\subsection{The Role of the Decadal Review}

The problems of the postdoc system are not limited to astronomy, nor
can they be solved overnight. One advantage that our field has,
however, is relatively small size and the importance that federal
funding plays within it. {\bf We recommend that the SSP}:

\parindent 0.5in

{\bf (D)} Re-evaluate postdoctoral and pre-tenure positions and recommend
funding changes to remove the ``arms race'' incentives of the current
system.

\parindent 0.25in

We hope that the SSP discusses a wide range of approaches to meeting
this challenge, such as: eliminating the plethora of federally funded
postdoctoral fellowships in favor of funding more diverse permanent
positions; completely reconceptualizing the postdoctoral process and
the transition from graduate school to permanent positions;
encouraging the creation of a more fluid workforce in which
early-career jobs can be held for longer periods of time and
transitions between positions flow naturally with project timescales
rather than in rigid (typically 3-5 year) timescales. It is worth
noting that the tenure system plays a fundamental role in shaping the
current postdoctoral system. Finally, in this topic, academic systems
outside of the United States can provide examples, both good and bad,
from which to learn.

\section{Retaining Excellent Astronomers}

The lack of adaptation in our professional institutions to social
change has dramatic effects on the retention of excellent astronomers,
with particular impact on underrepresented groups. Lengthy reports can
be (and have been) written on this topic; we will focus on one
example, the area of ``child-friendliness'' within astronomy and its
effect upon women in our field. We will treat it briefly and consider
again how the structure of our institutions affects the profession; we
hope that other position papers will deal with this, and the other
ways in which our field loses excellent astronomers, much more fully.

\subsection{Child-Friendliness}

In the past, most couples had one working spouse and one spouse who
performed virtually all childcare duties. In such situations, it is
viable for the working spouse to have an extremely time-demanding
job. The modern norm is for families to be dual-income, and
professional couples with children increasingly expect that both
partners will work hard, pursue a fulfilling career, and share in
childcare duties [13]. In this situation time-demanding and relatively
low-paying jobs are much more difficult to accommodate.

The increase in demand for careers that accommodate two-income
families has occurred more-or-less simultaneously with
the severe lengthening of the postdoc phase. All of the difficulties
mentioned in \S3 are particularly problematic for parents; moreover,
the postdoc stage usually happens at the exact age --- in the late
20's and early 30's --- in which most families are started. The
hardship of multiple relocations is especially trying for those with
long-term partners --- let alone those with long-term partners {\it
  and} children --- who therefore need to solve the ``two-body
problem'' not once but several times over the course of only a few
years. The reality of these concerns is well-established in our field
and others. In a survey of University of California graduate students
in all disciplines, 74\% of the male respondents and 84\% of the
female respondents reported being ``somewhat'' or ``very concerned''
about the family-friendliness of their career paths [14]. (Here, we
treat issues relating to family-friendliness as a superset of those
relating to child-friendliness.) Exacerbating the problem is the
difficulty of re-entering the field after any significant time away
from it, discouraging would-be parents from leaving the field
temporarily to care for young children.  In fact, even junior faculty
are uncomfortable with lessening their workload for
maternity/paternity.  According to the UC Faculty Work and Family
Survey [15], less than a third of eligible faculty used the
University's tenure ``clock stoppage'' option for new parents; of the
survey respondents who did not, a significant fraction ($\sim$30\%)
cited ``It might have hurt my career'' as a reason for not invoking
it.

\subsection{Impact of Child-Unfriendliness on Excellence}

While the difficulties of raising a family affect all academics, there
is no question that they disproportionately impact the careers of
women, regardless of their level of talent. In [14], 46\% of the
female respondents who began graduate school with the goal of becoming
faculty but shifted their goals cited ``issues related to children''
as a major factor, while only 21\% of the male respondents did. The
results of [15] make for sobering reading: in virtually every aspect,
female faculty find more of a tension between their careers and their
families than their male counterparts. {\bf We emphatically reject the
  notion that a prioritization of family life over career necessarily
  implies a lack of excellence in the field.} Improving the
representation of women in astronomy depends upon addressing the
child-friendliness of academic careers, though {\bf the status of
  women in astronomy depends on far more than this one
  factor}. Conversely, {\bf genuine efforts in this direction will
  make it a better field for all of its members, not just women.}

\subsection{Impact on Legitimacy of the Field}

We believe that the inequities alluded to in this section are all
serious and worthy of correction on their own merits. However, they
also have a damaging effect on the field by injuring its legitimacy in
the public eye. Legislators and other funders may question ---
justifiably --- whether they should direct spending towards a field
that is only sluggishly addressing its glaring inequalities [16,
  17]. This is especially true for a field which generally aspires to
be a meritocracy in which individuals succeed purely based on the
quality of their contributions. While astronomy is not the only
offender, other fields, notably biology, do a better job of retaining
excellent scientists [18, 19].

Significant federal effort is now directed towards rectifying gender
inequalities in STEM (Science, Technology, Engineering, and
Mathematics) professions. In [17], the National Academies stated that
in order to ``maintain scientific and engineering leadership amid
increasing economic and educational globalization, the United States
must aggressively pursue the innovative capacity of {\it all} of
its people,'' regardless of gender (emphasis original). In October
2008, Barack Obama responded to a query from the Association for Women
in Science with the following statement of policy [20]:
\begin{quote}
We will need to significantly increase our STEM
workforce, and to do that we will need to engage not just women and
minorities but also persons with disabilities, English language
learners, and students from low income families\ldots\ We also support
improved educational opportunities for all students, increased
responsibilities and accountability for those receiving federal
research funding, equitable enforcement of existing laws such as Title
IX, continuation and strengthening of programs aimed at broader
engagement in the STEM disciplines\ldots
\end{quote}
There has, in fact, been formal federal investigation into forcing
university science programs to begin addressing inequalities by
applying Title IX to them (\eg\ [21, 22]).

\subsection{The Role of the Decadal Review}

It will take concerted effort across many sectors of academia,
government, and society to enable us to attract and retain the best
astronomers from all demographics. Nevertheless, the Decadal Review is
an opportunity to begin implementing necessary policies. We {\bf
  suggest that the SSP}:

\parindent 0.5in

{\bf (E)} Mandate that job ads and offer letters at {\it all} levels in the
field include information on the hiring institution's family-friendly
policies.

{\bf (F)} Identify model programs that have demonstrated positive impacts on
the demographics and family-friendliness of astronomy and recommend
that funding be allocated for duplications, expansions, and
improvements of these programs.

{\bf (G)} Identify policies that help retain the most talented
astronomers and recommend that required implementation of these
policies be attached to federal funding. Study the examples of other
fields for lessons, both positive and negative.

{\bf (H)} Consider any changes to the postdoctoral system in the light
of the effect they will have on family-friendliness and the retention
of excellent astronomers.

\parindent 0.25in

We reiterate that {\bf many of the current challenges in astronomy
  careers are due to institutional structures than can be changed}.
Some ideas to initiate discussion are: establishing opt-out minimum
tenure ``clock stopping'' or parental leave policies; issuing
comprehensive employer childcare assistance standards. A community
approach to the enforcement of Title IX, should it be mandated, should
be discussed.

\section{Conclusion}

The Decadal Review provides an invaluable opportunity to affirm {\it
  and revise} our values as a community and set priorities
accordingly.  This process has been tremendously successful in
establishing support for instruments that have been the basis for
ground-breaking science: the Very Large Array, HST, and Spitzer were
all made possible in large part due to recommendations of past Decadal
Review committees.  The Decadal Review process has also won support in
Washington for the field of astronomy and indirectly led other fields
to establish their own similar review processes [23].

However, while past reviews have been extremely successful in shaping
the technologies used to pursue the next generation of science, less
attention has been directed towards properly training and maintaining
the astronomers who perform the science. Still less has been focused
on training the next generation of astronomy educators, public policy
experts, and project managers. While the needs of the profession and
the labor market have evolved significantly since the astronomy
Decadal Review process was established, the overall academic structure
of the field has remained largely unchanged.  Many of our current
practices are outmoded, resulting in misallocated resources and
attrition patterns that cause us to lose the contributions of
excellent scientists.

The failure of past Decadal Review processes to allocate sufficient
time and funding to revising these practices represents an
undervaluation of the field's human resources.  In a time of
economic downturn and budget shortfalls, it is in our best interest to
put stock in the ability of talented individuals to develop creative
new solutions to outstanding problems in our field, whether those
problems be in basic research, education, public outreach, or policy.
We urge the Subcommittee to develop strong, concrete recommendations
tied to funding which acknowledge and support the important role human
contributions make to the scientific endeavor.

\section*{References}

\begin{hangparas}{0.25in}{1}

[1] Davis, G. ``Doctors Without Orders'', 2005, {\it American 
  Scientist}, 93, 3 (supplement): \url{http://postdoc.sigmaxi.org/results/}

[2] Thronson, H.~A. 1991, {\it PASP}, 103, 90

[3] Metcalfe T. S. 2008, {\it PASP}, 120, 229

[4] Mulvey, P. J. \& Nicholson, S. 2008, American Institute of Physics
(AIP) Report, R-151.43:
\url{http://www.aip.org/statistics/trends/reports/ed.pdf}

[5] ``NSF Statistics on S\&E Doctorate Awards'',
\url{http://nsf.gov/statistics/nsf07305}

[6] ``First Triple Quasar Discovered at W. M. Keck Observatory'', 8
Jan 2007, {\it News and Outreach}, W. M. Keck Observatory:
\url{http://keckobservatory.org/index.php/news/first_triple_quasar_discovered_at_w._m._keck_observatory/}

[7] ``A Guide to the Astronomy Graduate Program'', 2007, University of
Arizona, Department of Astronomy and Steward Observatory (12 Mar
2009):
\url{http://www.as.arizona.edu/academic_program/graduate_program/graduate_academic_guide.html}

[8] Gilreath, J. A. \& Slater, T. F. 1994, {\it Phys. Educ.}, 29, 200

[9] Slater, T. ``Re: Training Grad Students as Teachers'', 11 Mar
2009, astrolrner Message Board - NASA CAE: Improving Astronomy
Education:
\url{http://tech.groups.yahoo.com/group/astrolrner/message/2450}

[10] Sage, L. ``Scope for improvement: Assisting astronomy's rising
stars'', 2001, {\it Nature}, 414, 4-5:
\url{http://www.nature.com/nature/journal/v414/n6864/full/nj6864-04a.html}

[11] Zimmerman, R. {\it The Universe in a Mirror: The Saga of the
  Hubble Space Telescope and the Visionaries Who Built It}, 2008,
Princeton University Press.

[12] Astronomy and Astrophysics Survey Committee, Commission on
Physical Sciences, Mathematics, and Applications, National Research
Council. ``Astronomy and Astrophysics in the New Millennium'', 2001,
The National Academies Press: 
\url{http://www.nap.edu/openbook.php?isbn=0309070317}

[13] Winkler, A. E. ``Earnings of husbands and wives in dual-earner
families'', 1998, {\it Monthly Labor Review}, 121, 42:
\url{http://www.bls.gov/opub/mlr/1998/04/art4full.pdf}

[14] Mason, M. A. \& Goulden, M. ``UC Doctoral Student Career Life
Survey'', 2006, The UC Faculty Family Friendly Edge:
\url{http://ucfamilyedge.berkeley.edu/Why%20Graduate%20Students%20Reject%20the%20Fast%20Track.pdf}

[15] Mason, M. A., Stacy, A., Goulden, M., Hoffman, C., and Frasch,
K. ``UC Faculty Family Friendly Edge Report'', 2005, The UC Faculty
Family Friendly Edge:
\url{http://ucfamilyedge.berkeley.edu/ucfamilyedge.pdf}

[16] Bagenal, F. ``The Leaky Pipeline for Women in Physics and
Astronomy'', June 2004, {\it STATUS: A Report On Women in Astronomy},
13-19: \url{http://www.aas.org/cswa/status/STATUS_Jun04sm.pdf}

[17] ``Beyond Bias and Barriers: Fulfilling the Potential of Women in
Academic Science and Engineering'', 2006, National Academies Report:
\url{http://www.nap.edu/catalog.php?record_id=11741#toc}

[18] ``Digest of Education Statistics'', Fall 2005, U.S. Department of
Education, National Center for Education Statistics, 2004--05
Integrated Postsecondary Education Data System (IPEDS):
\url{http://nces.ed.gov/programs/digest/d06/tables/dt06_258.asp}

[19] Leadley, J., Magrane, D., Lang, J., and Pham, T. ``Women in U.S.
Academic Medicine: Statistics and Benchmarking Report, 2007-08'',
2008, Association of American Medical Colleges (AAMC):
\url{http://www.aamc.org/members/wim/statistics/stats08/stats_report.pdf}

[20] ``Campaign Responses to Questions from The Association for Women
in Science \& The Society of Women Engineers'', 15 Oct 2008,
Association for Women in Science (AWIS):
\url{http://www.awis.org/documents/ObamaMcCainResponses.pdf}

[21] Mervis, J. ``Can Equality in Sports Be Repeated in the Lab?'',
2002, {\it Science}, 296, 356:
\url{http://www.sciencemag.org/cgi/content/full/298/5592/356}

[22] ``Gender Issues: Women's Participation in the Sciences Has
Increased, but Agencies Need to Do More to Ensure Compliance with
Title IX'', 22 Jul 2004, Government Accountability Office (GAO)
Report, GAO-04-639: \url{http://www.gao.gov/products/GAO-04-639}

[23] Brumfiel, G. ``Wishing for the Stars'', 2006, {\it Nature}, 443, 386

\end{hangparas}

\end{document}